\documentclass[reprint,twocolumn,superscriptaddress,nofootinbib]{revtex4-2}
\usepackage{amsmath,amssymb,amssymb,amsfonts,graphicx,float,dcolumn,bm,color,colortbl,multirow,fixltx2e,mathtools}
\usepackage[titletoc,title]{appendix}
\usepackage[dotinlabels]{titletoc}
\usepackage{cancel,braket,mathrsfs,comment}
\usepackage[colorlinks=True,citecolor=blue,linkcolor=blue,urlcolor=blue]{hyperref}
\usepackage{braket}
\usepackage[mathscr]{euscript}
\usepackage[dvipsnames]{xcolor}
\newcommand{\eq}[1]{\begin{align}#1\end{align}}

\newcommand{\PoP}{{\rm PoP}}

\begin{document}
\title{Projected ensemble in a system with conserved charges with local support}
\author{Sandipan Manna}
\email{sandipan.manna@students.iiserpune.ac.in}
\affiliation{Indian Institute of Science Education and Research, Pune 411008, India}

\author{Sthitadhi Roy}
\email{sthitadhi.roy@icts.res.in}
\affiliation{International Center for Theoretical Sciences, Tata Institute of Fundamental Research, Bangalore 560089, India}

\author{G. J. Sreejith}
\email{sreejith@acads.iiserpune.ac.in}
\affiliation{Indian Institute of Science Education and Research, Pune 411008, India}
\date{\today}
\begin{abstract}
The investigation of ergodicity or lack thereof in isolated quantum many-body systems has conventionally focused on the description of the reduced density matrices of local subsystems in the contexts of thermalization, integrability, and localization. Recent experimental capabilities to measure the full distribution of quantum states in Hilbert space and the emergence of specific state ensembles have extended this to questions of {\textit{deep thermalization}}, by introducing the notion of the {\textit{projected ensemble}} -- ensembles of pure states of a subsystem obtained by projective measurements on its complement. While previous work examined chaotic unitary circuits, Hamiltonian evolution, and systems with global conserved charges, we study the projected ensemble in systems
where there are an extensive number of conserved charges all of which have (quasi)local support. We employ a strongly disordered quantum spin chain which shows many-body localized dynamics over long timescales as well as the $\ell$-bit model, a phenomenological archetype of a many-body localized system, with the charges being $1$-local in the latter. 
In particular, we discuss the dependence of the projected ensemble on the measurement basis. Starting with random direct product states, we find that the projected ensemble constructed from time-evolved states converges to a Scrooge ensemble at late times and in the large system limit except when the measurement operator is close to the conserved charges. This is in contrast to systems with global conserved charges where the ensemble varies continuously with the measurement basis. We relate these observations to the emergence of Porter-Thomas distribution in the probability distribution of bitstring measurement probabilities.
\end{abstract}
\maketitle
\section{Introduction}
The study of dynamics of local observables in generic isolated quantum many-body systems show that their expectation values equilibrate to a thermal value due to entanglement between subsystems and their complements. The Eigenstate Thermalization Hypothesis (ETH)~\cite{deutsch1991quantum,srednicki1994chaos,rigol2008thermalisation,D_Alessio_2016} formalizes this and relates this to the nature of eigenstates of generic 
Hamiltonians  
- reduced density matrices of local subsystems in the eigenstates match thermal ensembles determined by the eigenstate's energy density.
 
While conventional experiments are able to measure expectation values of local observables, modern quantum devices can make projective measurements of multiple degrees of freedom simultaneously through techniques like single-atom resolved fluorescence~\cite{haller2015single} blurring the system-bath distinction~\cite{choi2023preparing,neill2016ergodic}. In such settings, the configuration of the entire system (e.g.~an array of Rydberg atoms \cite{ebadi2021quantum}/superconducting qubits~\cite{opremcak2021high}) is read out as a classical bitstring, thus capturing information at the level of individual degrees of freedom of the many body system ~\cite{shaw2024universal,bakr2009quantum,bernien2017probing,kaufman2016quantum}. Statistical information of such repeated measurements enable investigation beyond simple expectation values of local observables described by subsystem density matrices. 

These modern experimental capabilities have motivated the notion of a {\it projected ensemble}~\cite{cotler2023emergent,choi2023preparing} -- ensemble of pure states on a local subsystem obtained from the outcome of projective measurements on its complement with the latter associating Born rule probabilities to the members of the ensemble. This defines a natural one among all purifications~\cite{HUGHSTON199314} of the subsystem's reduced density matrix.
More explicitly,
given 
multiple copies of a pure state 
$\ket{\psi_{AB}}$ of a system partitioned into $A$ and $B$, the projected ensemble (PE) 
denoted as 
\begin{equation}
\mathcal{E}_{\rm PE}(\psi_{AB}) \equiv \{ p(b_i),~\ket{\psi_A(b_i)}\}_{b_i\in {\cal S}_B}^{\phantom\dagger}\,,
\label{eq:PEdef}
\end{equation}
is generated by complete set of projective measurements in $B$ on the copies, each yielding a pure state 
$\ket{\psi_A(b_i)}$ on subsystem $A$ with a probability given as $p(b_i) \equiv \langle \psi^{AB} | (\mathbb{I}_A \otimes |b_i\rangle \langle b_i)| \psi^{AB} \rangle$ where ${\cal S}_B$ is the set of all possible measurement outcomes in $B$ (for example, $b_i$ can be a bitstring for a system of qubits) and $\mathbb{I}_A$ is the identity operator acting on subsystem $A$.

Conventionally, the ensemble can be characterized by its density matrix
\begin{equation}
\rho_A = \sum_i p_i\ket{\psi_A(b_i)}\bra{\psi_A(b_i)}\equiv \sum_i p_i\rho_A(b_i)\,.
\end{equation}
and contains all information about $\mathcal{E}_{\rm PE}(\psi_{AB})$ that an external agent can possibly infer by measuring the pure states in the ensemble. The expectation value in the ensemble of a local observable $O_A$ is the first moment of the expectation values of $O_A$ in the states in the ensemble \eqref{eq:PEdef}, or equivalently $\braket{O_A} = \sum_ip_i\braket{O_A}_{\psi_A(b_i)}$ and is fully determined by the density matrix. 

Ensembles on the other hand contain more information in terms of the distributions of the states $\ket{\psi_A(b_i)}$ in the Hilbert space and can be characterized by higher moments of the distribution of states beyond the first moment. In PEs, the external agent can estimate the higher moments if the states $\psi_A(b_i)$ are supplemented with the measurement outcome data $b_i$.

This has given rise to the concept of {\it deep thermalization}~\cite{cotler2023emergent,harrow2023approximate} which is a statement not just about $\rho_A$ but regarding the entire ensemble \eqref{eq:PEdef} (and therefore arbitrary moments of it) approaching universal ensembles under generic quantum dynamics at late times.
For example, in systems without any conservation laws, 
the PE approaches the Haar ensemble~\cite{harrow2013church, zyczkowski2001induced} as has been demonstrated in experiments~\cite{choi2023preparing}.
In systems with global charge conserving dynamics, starting from an atypical state such as a direct product state, the PE relaxes to various limiting ensembles at late time~\cite{mark2024maximum}. These limiting ensembles maximize the entropy in the state distribution over the accessible Hilbert space while respecting the conserved charges and constraints of the dynamics, suggesting a generalization of the second law of thermodynamics. 

For example, under Hamiltonian dynamics (with energy conservation manifest) the PE approaches the Haar ensemble only when the initial state is at efffective infinite temperature whereas for finite temperatures it approaches the so-called Scrooge ensemble~\cite{jozsa1994lower,goldstein2006distribution}
if the measurements in $B$ are uncorrelated with the conserved charges. 
Correlation of measurement basis with the conserved charges changes the limiting ensemble from Scrooge~\cite{jozsa1994lower,goldstein2006distribution} to a generalized Scrooge ensemble (GSE)~\cite{chang2024deep} which is a convex combination of multiple Scrooge ensembles. 

PE and related ideas have been studied in a variety of systems. The non-interacting free-fermion system was found to approach an ensemble determined by just the conserved charges of initial state~\cite{lucas2023generalized}. A $U(1)$ charge conserving random unitary circuit was also investigated with focus on the effect of measurement basis~\cite{chang2024deep}. Studies have been carried out in constrained systems~\cite{bhore2023deep} as well. In these cases too, it is expected that the general ensemble should be described by a maximum-entropy ensemble on the accessible Hilbert space satisfying the conditions imposed by the conserved charges in the initial state and the effective constraints on dynamics, thus describing a quantum Hilbert space ergodicity. Deterministic driving was also found to lead to the Haar ensemble~\cite{pilatowsky2023complete}. In few specific cases such as the random unitary and dual-unitary circuits, PEs are analytically tractable with replica method~\cite{ippoliti2023dynamical,shrotriya2023nonlocality}.

While much of the work in this context so far has been on systems with no conservation laws or with global conserved charges, in this work, we focus on a system where the conserved charges have local support. And yet, for the late time ensemble of states to be described by universal ensembles, a necessary ingredient is that the systems should equilibrate, albeit to non-thermal ensembles. Many-body localized (MBL) systems~\cite{nandkishore2015many,abanin2017recent,abanin2019colloquium,sierant2024manybody} represent a class of models which satisfy both the above requirements as they posses an extensive number of locally conserved charges~\cite{serbyn2013local,huse2014phenomenology,imbrie2017local,ros2015integrals} and yet the dynamics shows dephasing resulting in approach to equilibrium~\cite{serbyn2014quantum,bardarson2012unbounded,serbyn2013universal}.

We study the limiting ensembles and their dependence on measurement basis for initial states with finite values of local conserved charges. We observe that in this case, based on the first three moments of the distributions, PEs approach a Scrooge ensemble for any measurement basis in thermodynamic limit except when the measurement basis has a large overlap with the conserved charges. 

In chaotic systems, the emergence of the universal ensembles (such as Haar or Scrooge) as the limiting cases for the projected ensembles are intimately connected to universal probability distributions of measurement-outcome probabilities, such as the Porter-Thomas distribution for the Haar ensemble. In our case also, we relate the  measurement-basis dependence of the projected ensemble to the probability distribution of measurement-outcome probabilities and their deviations from the Porter-Thomas distribution.

This article is organized as follows. In Sec.~\ref{PE_defn}, we give a brief review of projected ensemble and its construction. We also introduce Scrooge ensemble in this section. We numerically demonstrate the convergence of projected ensemble to Scrooge ensemble for a Floquet MBL model in Sec.~\ref{sec:floquet-ising}. The phenomenological $\ell$-bit model and PE constructed under time-evolved states generated with it is described in Sec.~\ref{l-bit model} where numerical results for the convergence of the PE to the Scrooge ensemble is also presented. The emergence of Porter-Thomas distribution in temporal ensemble is investigated in Sec.~\ref{sec:temp-ens-pop}. Subsequently, we derive an expression for moments of the PE constructed with time-evolved states under $\ell$-bit Hamiltonian in Sec.~\ref{Main: matrix_el_PE_Sc}. 
We conclude our discussion in Sec.~\ref{Conclusion} commenting on possible extensions of our result.

\section{Projected ensemble and Deep thermalization}
\label{PE_defn}
 Let $\ket{\psi_{AB}}\in \mathcal{H}_A \otimes \mathcal{H}_B$ be a pure state of a composite system of $L=L_A+L_B$ qubits (local Hilbert space dimension of 2). Given many copies of $\ket{\psi_{AB}}$, one can perform projective measurements on partition $B$ in a complete orthogonal basis $\{b_h\}_{h\in[1,2^{L_B}]}$ which can be implemented using projectors of the form $\Pi_{b_h}=\mathbb{I}_A\otimes |b_h\rangle \langle b_h|$.  Each measurement will produce a post-measurement pure state labeled by the measurement outcome $b_h$ in $B$:
\begin{align}
    |\psi^{b_h}\rangle = \frac{1}{\sqrt{p(b_h)}} \Pi_{b_h}\psi_{AB} = \frac{1}{\sqrt{p(b_h)}}|\psi_A({b_h})\rangle \otimes |b_h\rangle
\end{align}
where $p(b_h)=\langle \psi_{AB}| \Pi_{b_h} | \psi_{AB} \rangle $ is the probability of obtaining measurement outcome $b_h$ in $B$ and $|\psi_A({b_h})\rangle$ is the conditional state in $A$ given the outcome $b_h$. 
In this work, we will  consider the set of states $\{b_h\}$ to be the joint eigenbasis of local spin operators $\vec{\sigma}.\hat{n}$ on each site along some arbitrary direction $\hat{n}$.

The projected ensemble is constructed with the pure states obtained from measurements and their corresponding probabilities, $\mathcal{E}_{\rm PE}=\{p(b_h),|\psi_A(b_h) \rangle\}$, as in Eq.~\ref{eq:PEdef}. An external agent, if provided with the closely related ensemble $\tilde{\mathcal{E}}=\{p(b_h),|\psi^{b_h} \rangle\}$, can estimate the distribution of states in $\mathcal{E}_{\rm PE}$ from expectation values of non-local observables in $\tilde{\mathcal{E}}$.
The protocol for the construction of the ensemble ensures that in the ensemble, two different states in $A$ are not associated with the same $b_h$ i.e. $\psi_A({b_h})\neq \psi_A({b_h'})$ implies $b_h\neq b_h'$. The label $b_h$ then allows locating repeated occurrances of the same state $|\psi_A({b_h}) \rangle$ and makes it possible to estimate conditional expectation values $\langle \hat{O} \rangle_{\psi_A({b_i})}$ of observables for each state in the ensemble.  The \textit{k}-th moment of the distribution of expectation values of $\hat{O}$ is $\sum_i p(b_i) \langle \hat{O} \rangle_{\psi_A^{b_i}}^k $. Moments of all observables are directly related to \textit{k}-th moment of the ensemble defined as follows.
\begin{align}
\label{moment definition}
\rho^{(k)}_A=\sum_{b_i\in {\cal S}_B} p(b_i) (|\psi_A({b_i}) \rangle  \langle \psi_A({b_i}) |)^{\otimes k }  
\end{align}
These moments can be used to characterize and compare distribution of states in different ensembles.

For a random state in the Hilbert space, the PE will converge to a Haar ensemble in the $A$ subsystem. This holds true even when $\psi_{AB}$ is obtained by time evolution of an effectively infinite temperature state of a chaotic Hamiltonian\cite{cotler2023emergent}. The proximity of two ensembles can be gauged by the trace distance between their $k$-th moments,
\eq{
\Delta^{(k)}(\mathcal{E},\mathcal{E}')=\frac{1}{2}{\rm Tr}\left[\sqrt{\left(\rho^{(k)}_\mathcal{E} - \rho^{(k)}_{\mathcal{E}'}\right)^\dagger\left(\rho^{(k)}_\mathcal{E} - \rho^{(k)}_{\mathcal{E}'}\right)}\right]\,.
\label{Tr_dist_formula}
}

The Haar distribution for $A$ can be written as a measure over normalized quantum states $|\psi\rangle$ as below,
\begin{equation}
P_{\text{Haar}}(\psi) = \int^{\infty}_{-\infty} \delta\left(\psi- \frac{\tilde{\Psi}}{\|\tilde{\Psi}\|}\right)\prod_{i=1}^D\frac{D}{\pi}\exp[-D|\tilde{\Psi}_{i}|^2] d^2\tilde{\Psi}_i
\end{equation}
where $D$ is the Hilbert space dimension of $A$.

For systems with global conserved charges $\{Q\}$, the reduced density matrix $\rho_A$ equilibrates to $\exp[{-\sum_Q \lambda_Q Q_A}]$ and the the limiting ensemble for the PE is found to approach the Scrooge ensemble\cite{chang2024deep,cotler2023emergent} if the measurement basis $\{b_h\}$ does not reveal any information about the density of conserved charges (e.g. energy) in $B$ i.e have 0 overlap with the the eigenstates of conserved charges (energy). The measurement in the joint eigenbasis of $\{\sigma^{x}_i\}_{i\in B}$ for a system with a globally conserved $U(1)$ charge ${Q=\sum_{i=1}^L\sigma^z_i}$, would be such an example.

Scrooge ensemble ($\mathcal{E}_{\text{Scrooge}}$) can be constructed as a distortion of the Haar ensemble such that it reproduces a given density matrix $\rho$ as its first moment. Formally, we can represent it as the following,
\begin{align}
\mathcal{E}_{\text{Scrooge}}(\rho) = \left\{D\langle \psi | \rho | \psi \rangle, \frac{\sqrt{\rho}|\psi\rangle}{\sqrt{\langle \psi | \rho | \psi \rangle}}\right\}_{\psi \in \mathcal{E}_{\text{Haar}}}
\label{eq:Scrooge-def}
\end{align}
Given a $\rho$, the Scrooge ensemble \cite{jozsa1994lower,goldstein2006distribution}  is the unique ensemble such that it has the minimum accessible information as shown in Ref.~\cite{jozsa1994lower}, thus manifesting a maximal entropy principle\cite{mark2024maximum} in the full Hilbert space. The moments of the Scrooge ensemble are given by 
\begin{align}
\label{rho k first expression}
    \rho_{\text{Scr}}^{(k)}=\int d\psi~ \left(\frac{\sqrt{\rho}|\psi\rangle \langle \psi|\sqrt{\rho}}{|\sqrt{\rho}|\psi\rangle|^2}\right)^{\otimes k} \langle \psi | \rho | \psi \rangle P_{\text{Haar}}(\psi)
\end{align}

If the measurement basis reveals partial information about the conserved charges in $B$, it is conjectured that the PE at late times, converges to a weighted sum of Scrooge ensembles,
referred to as the generalized Scrooge ensemble (GSE)~\cite{mark2024maximum,chang2024deep}. Consider a global conserved charge and a measurement operator on each site that has a finite overlap with conserved charge. There is a probability of obtaining any given outcome $b_h$ which we denote as $p_d(b_h)$ and an associated reduced density matrix $\overline{\rho}_A(b_h)$. The charge in $A$ and therefore its density matrix is constrained by the charge in $B$ which is in turn constrained by the measurement outcome $b_h$. The $k$-th moment of the GSE is then given,
\begin{align}
\rho^{(k)}_{\text{GSE}}=\sum_{b_h} p_d(b_h) \rho^{(k),b_h}_{\text{Scr}}\,,\label{eq:generalizedScrooge}
\end{align}
where $\rho^{(k),b_h}_{\text{Scr}}$ is the $k$-th moment of the Scrooge ensemble corresponding to the density matrix $\overline{\rho}_A(b_h)$. For the special case where the outcome $b_h$ reveals no information about the conserved charge in $B$(and therefore no information about the conserved charge in $A$), the steady state density matrix $\overline{\rho}_A(b_h)$ is independent of $b_h$, resulting in Eq.~\ref{eq:generalizedScrooge} reducing to a single Scrooge ensemble.

If the independent set of conserved charges, instead, are locally supported, information about those in subsystem $B$ reveal no information about the charges in $A$. As a result, the steady state density matrix in $A$ conditional on the measured outcome in $B$ is independent of measured outcome. 
One expects that the projected ensemble of states in $A$ approaches, at late times, the same maximum entropy ensemble irrespective of the outcome in $B$, and therefore by the same arguments as above the unconditioned distribution approaches the Scrooge ensemble ${\cal E}_{\rm Scrooge}(\rho_{A,\infty})$ (see Eq.~\ref{eq:Scrooge-def}) with $\rho_{A,\infty}$ being the steady state reduced density of matrix of $A$ as $t\to\infty$. In the following we numerically demonstrate and semi-analytically prove for a specific model, that this is indeed the case -- this constitutes the central result of this work.

\section{Numerical results for a disordered Floquet spin chain \label{sec:floquet-ising}}
In this section, we establish numerically the phenomenology using a disordered Floquet spin-1/2 chain described by the  Floquet unitary $U_F$ given by
\eq{
    \begin{split}
    U_F &= \exp[-i\tau H_X]\,\exp[-i\tau H_Z]~~~{\rm with}\,,\\
        H_X &=g\Gamma\sum_{i=1}^L \sigma^x_i\,,\\
        H_Z &= \sum_{i=1}^{L-1}\sigma^z_i\sigma^z_{i+1} + \sum_{i=1}^{L}(h+g\sqrt{1-\Gamma^2}\epsilon_i)\sigma^z_i]\,,
    \end{split}
    \label{eq:UF}
}
where $\{\sigma_i^\mu\}$ is the set of Pauli matrices representing the spins-1/2 and $\epsilon_i\sim \mathcal{N}(0,1)$ are independent random numbers drawn from a standard Normal distribution.
The rationale behind choosing this model is twofold.
First, the model has been shown~\cite{zhang2016floquet} to exhibit MBL-like dynamics for numerically accessible system sizes over very long timescales, robustly, in the sense that the MBL regime has a finite extent in the parameter space. In particular for the choice of parameters, $g = 0.9045$, $h = 0.809$, and $\tau = 0.8$, there is a putative many-body localization transition at $\Gamma_c\approx 0.3$ with the model in an ergodic phase for $\Gamma>\Gamma_c$ and in a MBL regime for $\Gamma<\Gamma_c$. 
Second, the model has no global symmetries and in the MBL regime, the only conserved charges are the emergent local integrals of motion~\cite{serbyn2013local,huse2014phenomenology,ros2015integrals,imbrie2017local}. The model therefore allows us to understand cleanly the fate of the PE in the presence of locally supported conserved charges without any global conservation laws contaminating the results. 

We consider initial states of a direct product form
\begin{equation}
    \label{initial direct product state}
    |\psi_0 \rangle = \bigotimes\limits_{i=1}^{L} \left(\cos \frac{\theta_i}{2} \ket{\uparrow}_i + e^{\imath \phi_i} \sin \frac{\theta_i}{2} \ket{\downarrow}_i \right)\,,
\end{equation}
where $\ket{\uparrow}_i$($\ket{\downarrow}_i$) denote the state of the spin at site $i$ polarized along the positive(negative) $z$-axis.
The form in Eq.~\ref{initial direct product state} implies that initially the states of the individual spins are picked uniformly from the Bloch sphere with $\theta$ and $\phi$ (denoting the polar and azimuthal angles) are sampled accordingly.
Each initial state is therefore parameterized by a vector ${\bm \theta}\equiv\{\theta_1,\theta_2,...,\theta_L\}$ and ${\bm \phi}\equiv\{\phi_1,\phi_2,...,\phi_L\}$. 
We consider the subsystem $A$ to be the first $L_A$ sites of the chain.

As the explicit form of the locally conserved charge operators are unavailable, we directly construct the moments of the Scrooge ensemble as follows. For a given initial state the reduced density matrix of subsystem $A$ in the limit of infinite time is given by 
\eq{
\rho_{A,\infty} = \sum_{\omega}\left[{\rm Tr}_B \ket{\omega}\bra{\omega}\right]|\braket{\psi_0|\omega}|^2\,,
}
where $\{\ket{\omega}\}$ denotes the set of eigenstates of the Floquet unitary $U_F$ in Eq.~\ref{eq:UF}. With the $\rho_{A,\infty}$ at hand, and its eigenvectors and eigenvalues denoted as $\{\lambda_m,\ket{m}\}$, one can obtain  the $k$-th moment of the corresponding Scrooge ensemble as~\cite{mark2024maximum}
\eq{
\rho_{\rm Scr}^{(k)} = \sum_{{\bm m}}\sum_{\sigma\in S_k}\rho_{{\rm Scr},{\bm m}}^{(k)}\ket{{\bm m}}\bra{\sigma(\bm m)}\,,
}
where $\ket{{\bm m}} = \ket{m_1,m_2,\cdots,m_k}$ is a state in the $k$-replicated Hilbert space, $\ket{\sigma(\bm m)}$ is permutation of $\ket{{\bm m}}$, and the matrix element
\eq{
\rho_{{\rm Scr},{\bm m}}^{(k)} = \left(\prod_{m}\frac{1}{\lambda_m}\right)\frac{\partial^k \Lambda_k}{\partial\mu_{m_1}\cdots\partial\mu_{m_k}}\bigg\vert_{\mu_{i}=\lambda^{-1}_{i}}\,,
}
where
\eq{
\Lambda_k(\mu_1,\cdots) = \sum_{j}\frac{\mu_{j}^{k-2}\ln\mu_j}{\prod_{i\neq j}(\mu_j-\mu_i)}\,.
}

\begin{figure}
\includegraphics[width=\linewidth]{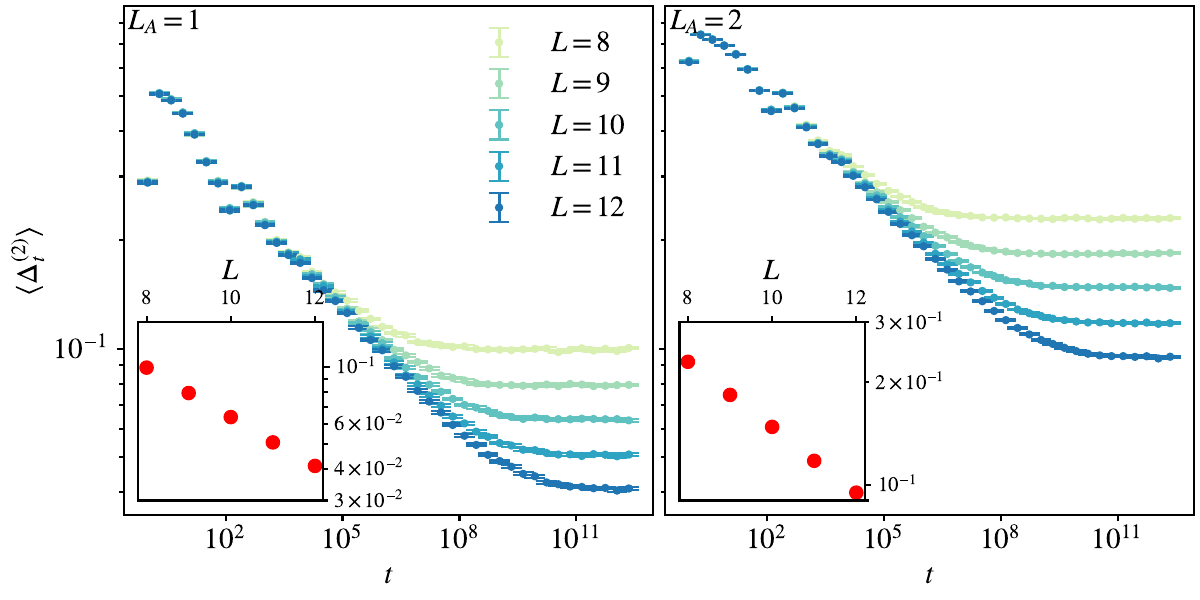}
\caption{Distance between the second moments of the Scrooge ensemble and the PE (see Eq.~\ref{eq:Delta-k-t} for definition) for the disordered Floquet spin chain (described by Eq.~\ref{eq:UF}) in the MBL regime with $\Gamma=0.15$. The left and right panels correspond to $L_A=1$ and $2$ respectively. The insets show the exponential decay of $\Delta_t^{(k)}$ with $L$ in the limit of $t\to\infty$. The trace distances are averaged over 500 disorder realizations and over 10 random initial states for each realization.
}
\label{fig:fim}
\end{figure}

We compare the so-obtained $k^{\rm th}$ moments of the Scrooge ensemble with the $k^{\rm th}$ moments of the projected ensemble, ${\cal E}_{\rm PE}(\psi_t)$ obtained from the time-evolved state $\ket{\psi_t} = U_F^t\ket{\psi_0}$.
In particular, we compute the trace distance defined in Eq.~\ref{Tr_dist_formula},
\eq{
\Delta^{(k)}_t\equiv \Delta^{(k)}\left({\cal E}_{\rm Scrooge}(\rho_{A,\infty}),{\cal E}_{\rm PE}(\psi_t)\right)\,,
\label{eq:Delta-k-t}
}
and study its behavior with $t$ and $L$ for fixed $L_A$.
The results for $k=2$ are shown in Fig.~\ref{fig:fim} where the measurement operators on $B$ are the $\sigma^x_i$ operators.
Note however, that the qualitative features of the results in this case are expected to remain the same for any generic set of single-site projective measurements on $B$ as the local conserved charges on $B$ can point along arbitrary directions and {can have a support with size greater than $1$} depending on the specific disorder realization of $U_F$.
The results in Fig.~\ref{fig:fim} show that $\Delta_t^{(2)}$ decays with $t$ and saturates to a values which decays exponentially in $L$. 
This provides strong hints towards the fact in the limit of $L\to\infty$ and $t\to\infty$, the PE is the same as the Scrooge ensemble corresponding to $\rho_{A,\infty}$. In the following sections, we show analytically that this is indeed the case using a phenomenological model of MBL.

\section{Phenomenological MBL model}	
\label{l-bit model}

As an effective, phenomenological model of MBL at strong disorder we consider the $\ell$-bit model~\cite{serbyn2013local,huse2014phenomenology,ros2015integrals,imbrie2017local}. The Hamiltonian for the model is given by
\begin{align}
    \label{l-bit Hamiltonian}
    H_{\ell{\text{-bit}}}=\sum_{i} J_i \sigma_i^z + \sum_{i<j} J_{ij} \sigma_i^z \sigma_j^z + \sum_{i<j<k}J_{ijk}\sigma_i^z \sigma_j^z \sigma_k^z\,,
\end{align}
with the couplings taken as random and decaying exponentially between the spins in their support as
\begin{align}
\label{couplin l-bit}
    J_i{=}u_i,\,\,J_{ij}{=}u_{ij}e ^ {-(j-i)/\xi},\,\,J_{ijk}{=}u_{ijk} e ^ {-(k-i)/\xi}\,,
\end{align} 
where $u$'s are picked from uniform distribution over $[-W,W]$, and $\xi$ is an effective localization length scale.
Throughout this work we take $W=1$ and time is always in units of $W^{-1}$, and $\xi$
is taken to be $2$ in our work. The model has strictly 1-local conserved charges $\{\sigma_i^z | 1\leq i\leq L \}$. In this case also we consider initial states of the form in Eq.~\ref{initial direct product state}.

\subsection{Approach of the projected ensemble to the Scrooge ensemble in time }

At long times, dephasing due to the interaction terms takes the reduced density matrix $\rho_{A}$ to a diagonal form in the energy eigenbasis which for the $\ell$-bit model (Eq.~\ref{l-bit Hamiltonian}) is just the $\sigma^z$-product state basis,
\begin{equation}
\rho_{A,\infty} \to \bigotimes\limits_{i=1}^{L_A} \rho^\infty_i\; {\text{ with }} \rho^\infty_i = {\rm diag}[\cos^2\frac{\theta_i}{2},\sin^2\frac{\theta_i}{2}]\,.
\label{eq:rhoA-inf}
\end{equation}
To construct the PE, as in the previous section, we time-evolve the initial state with the $\ell$-bit Hamiltonian (Eq.~\ref{l-bit Hamiltonian}), $\ket{\psi_t} = e^{-\imath H_{\ell{\text{-bit}}}t}\ket{\psi_0}$\, and perform measurements in $B$ subsystem of observable
$\otimes_{i=L_A+1}^L \sigma^n_i$ where $\sigma^n_i\equiv {\bm \sigma}_i\cdot{\hat{n}}$ is the Pauli spin operator on site $i$ in the direction $\hat{n}$. 
The latter is parametrized by the polar angle $\alpha\in[0,\pi]$ that $\hat{n}$ makes with the $z$-axis. The results are independent of the azimuthal angle defining $\hat{n}$ and therefore we choose it to be $0$ throughout.
The bitstrings $b_h\in(0,1)^{\otimes L_B}$ obtained from the measurement at an angle $\alpha$ correspond to states (in terms of the computational basis states $\ket{\uparrow}$ and $\ket{\downarrow}$ on each site),
\eq{
\begin{split}
\ket{0}&\equiv \cos\frac{\alpha}{2} \ket{\uparrow}{+}\sin\frac{\alpha}{2} \ket{\downarrow}\,,\\
\ket{1}& \equiv\sin\frac{\alpha}{2} \ket{\uparrow}{-}\cos\frac{\alpha}{2} \ket{\downarrow}\,.
\end{split}
\label{measurement basis defn}
}
Note that the limit of $\alpha=0$ is pathological as in that case the measurements coincide completely with the conserved charges. For much of what we discuss below we consider $\alpha\neq 0$.

\begin{figure}[!t]
\centering
\includegraphics[width=\linewidth]{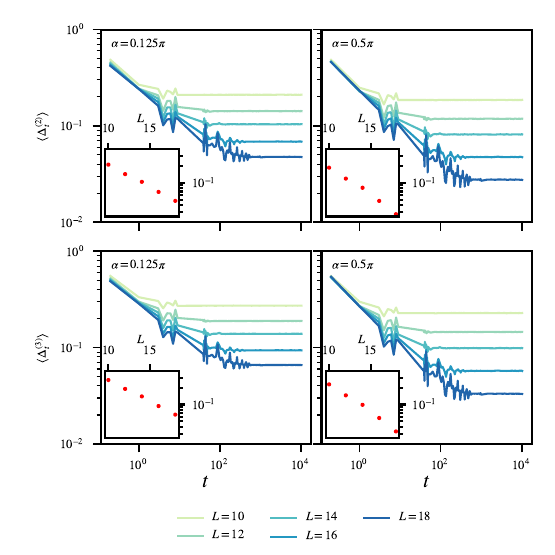}
\caption{Distance between the second and third moments of the PE, $\braket{\Delta_t^{(2)}}$ (top) and $\braket{\Delta_t^{(3)}}$ (bottom), from those of the Scrooge ensemble for two measurement angles, $\alpha=\pi/8$ (left) and $\alpha=\pi/2$ (right) for the $\ell$-bit Hamiltonian (Eq.~\ref{l-bit Hamiltonian}). For both cases, at long time, $\braket{\Delta_t^{(k)}}$ saturates to an asymptotic value which goes down exponentially with increasing $L$ (see insets). For these plots, $L_A=2$ and the trace distances are averaged over 500 disorder realizations and over 10 random initial states for each realization.}
\label{trace_dist_LA_2_both_distances}
\end{figure}

From the infinite-time reduced density matrix of $A$ in Eq.~\ref{eq:rhoA-inf}, the moments of the Scrooge ensemble can also be constructed directly as in the previous section and the distance of the PE from the Scrooge ensemble as a function of time can be studied using $\Delta_t^{(k)}$ defined in Eq.~\ref{eq:Delta-k-t},  as shown in Fig.~\ref{trace_dist_LA_2_both_distances} for $k=2$ and $k=3$. The results are qualitatively similar to those of the disordered Floquet Ising spin chain (see Fig.~\ref{fig:fim}) -- the trace distance of the moments of the PE and the Scrooge ensemble decay as a power-law in $t$, and for a finite $L$ saturate to a value which itself decays exponentially with $L$.
The power law exponent for decay of $\langle \Delta_t^{(3)}\rangle$ with time remains same as that of $\langle \Delta_t^{(2)}\rangle$ (within errorbar) and depends on $\alpha$. For the coupling strength we studied here,$\langle \Delta_t^{(2)}\rangle \propto t^{-0.33\pm0.06}$ for $\alpha=\pi/2$ and $\langle \Delta_t^{(2)}\rangle \propto t^{-0.24\pm0.05}$ for $\alpha=\pi/8$.  

\subsection{Dependence of the projected ensemble on the measurement angle}

In Fig.~\ref{trace_dist_LA_2_both_distances}, the results were shown only for two measurement angles $\alpha$. We next show results for how the moments of the PE depend on $\alpha$. 

\begin{figure}[]
\centering
\includegraphics[width=\linewidth]{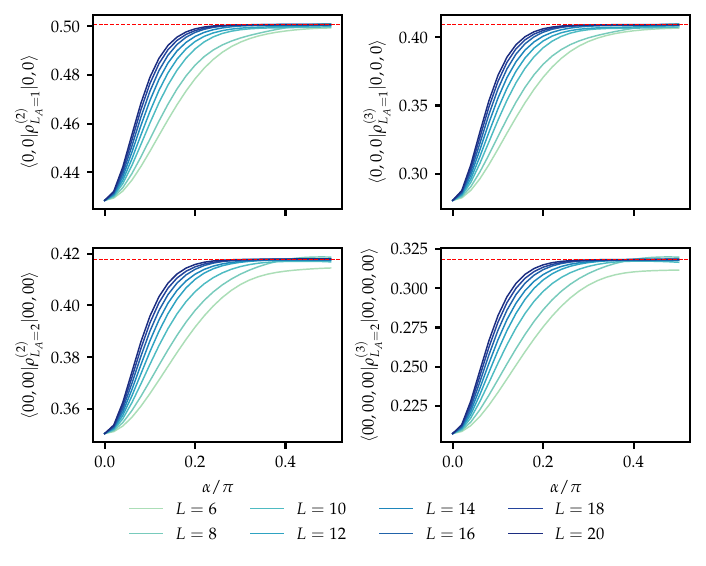}
\caption{Time averaged matrix elements of higher moments for different system sizes. (Top) Left panel shows $\langle 0,0 \vert \rho^{(2)}_{L_A=1} \vert 0,0 \rangle$ and right panels shows $\langle 0,0,0 \vert \rho^{(3)}_{L_A=1} \vert 0,0,0 \rangle$ where $\rho^{(2)}_{L_A=1},\rho^{(3)}_{L_A=1}$ are moments for $L_A$=1. (Bottom) Left panel shows $\langle 00,00 \vert \rho^{(2)}_{L_A=2} \vert 00,00\rangle$ and right panel shows $\langle 00,00,00 \vert \rho^{(3)}_{L_A=2} \vert 00,00,00 \rangle$ where $\rho^{(2)}_{L_A=2},\rho^{(3)}_{L_A=2}$ are moments for $L_A$=2. The red dotted lines are the corresponding Scrooge values. The initial state is a direct product state (Eq.~\ref{initial direct product state}) with $\theta_1=2\pi/5$ and $\theta_2=\pi/5$ (for $L_A=2)$.  To reduce fluctuation about mean value, we average the moments calculated at $90$ different times  at regular intervals in the time window $Wt=8000$ to $10000$. Legend shows the $L$ value corresponding to different colors. 
\label{fig:L_alpha_dependence_mat_elements}}
\end{figure}

\paragraph*{$\alpha$ dependence of the matrix elements:}
 First, we look at the asymptotic values of matrix elements in the moments of PE as a function of measurement angle $\alpha$. Representative results for matrix elements are shown in Fig.~\ref{fig:L_alpha_dependence_mat_elements}, for both $L_A=1$ and $2$.
The results show that individual matrix elements approach the Scrooge distribution's moments (red dotted lines) with increasing $L$ if the measurement is not aligned with the conserved charges. This qualitative trend appears to hold for all such matrix elements, which suggests that the PE approaches asymptotically the Scrooge ensemble in the limit of $L_B\to\infty$ for  $\alpha\gtrsim 0.1\pi$. These results on finite size effects and $\alpha$ dependence are in consonance with further independent numerical results in Sec.~\ref{sec:temp-ens-pop}. For $\alpha=0$, that is when the measurement is perfectly aligned with the conserved charges, the PE does not converge to Scrooge ensemble and shows no system size dependence. 
\begin{figure}[]
\centering
\includegraphics[width=\linewidth]{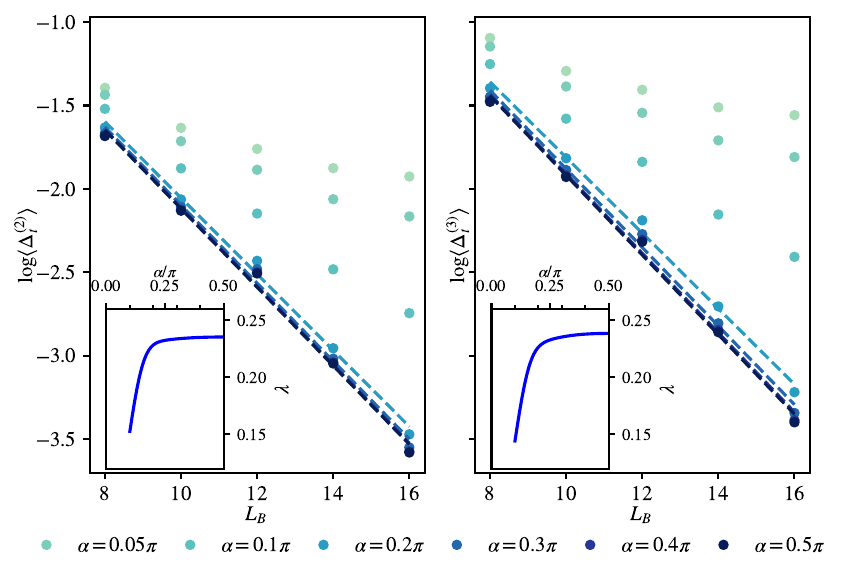}
\caption{Asymptotic trace distance between Scrooge ensemble and PE as a function of $L$ for $L_A=2$. Different colors denote different measurement angles $\alpha$ between $0.05\pi$ and $\pi/2$ (legend). The two panels show the trace distances for second (left) and third (right) moments.
Insets shows the decay rate $\lambda$ (estimated by fitting the trace distance to $e^{-\lambda L}$) as a function of measurement angles. The slope ($\lambda$) obtained from exponential fit for $\alpha > 0.1 \pi$. For other values of $\alpha$, the trend deviates substantially from an exponential within the range of $L$ studied.
The trace distances are averaged over 500 disorder realizations and over 10 random initial states at time $Wt=10000$.
\label{fig:asymptotic_trDist_L_slope_LA_2}}
\end{figure}

\paragraph*{$\alpha$ dependence of the trace distance:}
Finally, in Fig.~\ref{fig:asymptotic_trDist_L_slope_LA_2}, we present the asymptotic trace distance as a function of $L$. For  $\alpha\gtrsim \pi/8$, the trace distance (for both $\Delta^{(2)}$ and $\Delta^{(3)}$) between PE and Scrooge ensemble decays exponentially with $L$ as $\sim \exp(-\lambda L)$ with $\lambda\approx 0.23$. Interestingly, the same exponential decay rate was also found in the Floquet MBL model (Fig.~\ref{fig:fim}). For $\alpha$ close to $0$, the decay is slower and deviates from exponential form for the range of $L$ studied. This is consistent with the deviation of the matrix elements of the PE from those of the Scrooge ensemble for the available system sizes. 

To summarize, the numerical results presented in this section point towards the fact that for the $\ell$-bit model also, the PE at long times and large system sizes approaches the Scrooge ensemble unless the measurement basis closely aligns with the conserved charges. For $\alpha\gtrsim 0.1\pi$, scaling of the numerically obtained moments in the finite systems suggest that the trace distance vanish in the thermodynamic limit.
{In the following we will demonstrate this by showing semi-analytically, using empirical results from the probabilities of measuring bitstrings, that the matrix elements of the $k^{\rm th}$ moment of the PE at $L,t\to\infty$ are identical to those of the Scrooge ensemble.}

\section{Temporal ensemble and probability distribution of the probabilities (${\mathbf{PoP}}$) of bitstrings  \label{sec:temp-ens-pop}}

\begin{figure}[h!]
    \includegraphics[width=\linewidth]{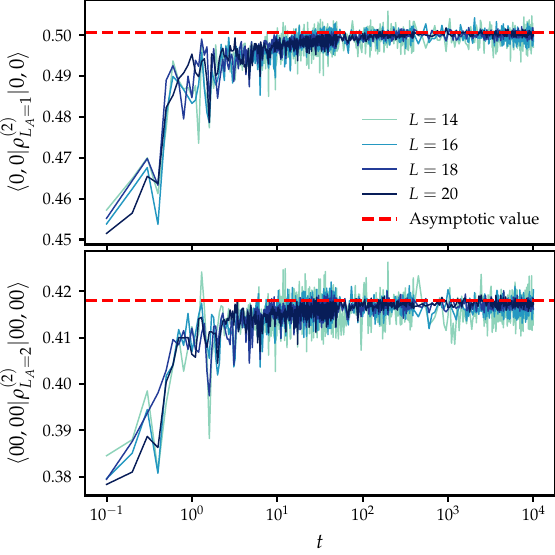}
    \caption{\label{fig:timeevo_single_state_alpha_0p25_mom2}
        The matrix elements in second moments of PE saturate at long timescale to their asymptotic value at measurement along an axis with $(\alpha) = \pi/4$ for both $L_A=1$ (Top) and $2$ (Bottom). The fluctuation about the asymptotic value (red dotted line) gets suppressed with increasing $L$. The initial state is a direct product state (eq. \ref{initial direct product state}) with $\theta_1=2\pi/5$ and $\theta_2=\pi/5$ (for $L_A=2).$
}
\end{figure}

A key observation that helps in computing the matrix elements of the higher-moments of the PE at late times is that they approach steady values with small temporal fluctuations that, crucially, decrease with increasing $L_B$. 
Representative numerical evidence for this is shown in Fig.~\ref{fig:timeevo_single_state_alpha_0p25_mom2} where it is seen that the matrix element of $\rho_A^{(2)}(t)$ fluctuate less around their steady-state values for larger systems sizes.
Other matrix elements of $\rho_A^{(2)}(t)$ and also matrix elements for other values of $k$ and $\alpha$ that we studied show similar qualitative behavior. 
This key empirical observation suggests that we can understand the steady-state values by performing an average of the matrix elements over time. 
This motivates the definition of a {\it{temporal ensemble}} of states consisting of the states sampled from the time-evolution trajectory of an initial state under a given Hamiltonian. Of particular significance will be understanding the probability distribution of the probabilities (PoP) of measuring individual bitstrings $b \in \{b_h\}$ of the entire system in the temporal ensemble of late-time states.

The temporal ensemble is defined as the collection of states $\{\ket{\psi_t}\}$ occurring in the quantum trajectory of the initial direct product state $\ket{\psi_0}$ (taken as in Eq.~\ref{initial direct product state}) over a suitably large temporal window at late times. Functionally, this can be taken to be the ensemble formed by sampling the states at times $t\in [\tau_1,\tau_2]$ at regular intervals of length $\delta t$:
\begin{equation}
\mathcal{E}_{\text{temp}}=\left\{\frac{\delta t}{\tau_2-\tau_1},e^{-\imath Ht}\ket{\psi_0}\right\}_{t\in[\tau_1,\tau_2]}\,.
\label{eq:TE-def}
\end{equation}

\begin{figure}[]
\centering
\includegraphics[width=\linewidth]{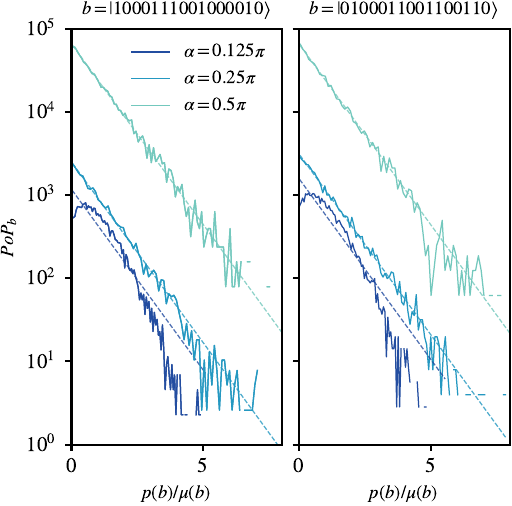}
\caption{Distribution of probabilities of  measurement bitstrings in the temporal ensemble, $\mathcal{E}_{\text{temp}}$, defined in Eq.~\ref{eq:TE-def}. ${\cal E}_{\rm temp}$ is constructed with 10000 states sampled in the time interval, $8000-10000$ (in unit of $W^{-1}$) with uniform spacing. Results are shown for three different measurement angles, $\alpha$, and the dotted lines show the Porter-Thomas distributions (defined in Eq.~\ref{eq:PT-def}) with the means obtained using Eq.~\ref{mu formula}. The two panels show two representative bitstrings (written above the figure) for a system with $L=16$.}
\label{fig:single_bitstring_ditribution_temporal}
\end{figure}

Upon measuring $\{\sigma^n_i\}_{i=1\dots L}$ at every site of the system, (recall $\sigma_i^n = {\bm \sigma}_i\cdot{\hat{n}}$ and $\hat n$ is parametrized by the polar angle $\alpha$ ) on the state $\ket{\psi_t}$, the probability of getting a bitstring $b$ is $p(b) = |\braket{\psi_t|b}|^2$.
Over the ensemble of states in the temporal ensemble (Eq.~\ref{eq:TE-def}) there is a probability distribution of the probability $p(b)$, and hence the name probability of probability (PoP). We denote the PoP of bitstring $b$ as $\PoP_b$. This is also the distribution of expectation values of the projector into a bitstring $b$ for states in the temporal ensemble.

Figure~\ref{fig:single_bitstring_ditribution_temporal} shows $\PoP_{b}$ over the temporal ensemble for two specific bitstrings (which are taken as representative examples) for three different values of $\alpha$, for a fixed initial state and disorder realization.

We find that the $\PoP_b$ is well-described by the Porter-Thomas distribution\footnote{Comparison of $\PoP$ with Porter-Thomas distribution is meaningful only for $\alpha\neq 0$. For the pathological case of $\alpha=0$, $\ket{b}$ are $\sigma^z$-product states and therefore coincide with the eigenstates of the $\ell$-bit Hamiltonian. As a result $p(b)$ for a given $b$ is a constant over the temporal ensemble, given by $|\braket{\psi_0|b}|^2$, and the PoP is $\delta$-function ditributed, $PoP_b(p) = \delta(p-|\braket{\psi_0|b}|^2)$.}
\begin{align}
PT_{b}(p)&=\frac{1}{\mu_b}e^{-\frac{p}{\mu_b}}\,,
\label{eq:PT-def}
\end{align}
where $\mu_b$ is the average of $p(b)$ over the temporal ensemble. 
In Fig. \ref{fig:single_bitstring_ditribution_temporal}, the PoPs are shown by the solid lines whereas the Porter Thomas distributions with the corresponding means are depicted by the dashed lines, with the two showing good agreement between each other. {The agreement is worse for $\alpha$ close to zero, which we will argue shortly is due to finite-size effects.}

The mean $\mu_b$ depends on the bitstring $b$, the initial state, the angle $\alpha$ parameterizing the measurement basis as 
\begin{align}
\label{mu formula}
\mu_b(\alpha,\psi_0)=\sum_{z_\nu}|\langle\psi_0|z_\nu\rangle|^2 |\langle b|z_\nu \rangle|^2\,,
\end{align}
where $\{\ket{z_\nu}\}$ are the eigenstates of the Hamiltonian, which for the $\ell$-bit Hamiltonian (Eq.~\ref{l-bit Hamiltonian}) are also $\sigma^z$-product states.
For $\alpha=\pi/2$, the bitstrings $\ket{b}$ are $\sigma^x$-product states such that $|\braket{b|z_\nu}|^2=D^{-1}$, with $D=2^{L}$ the Hilbert-space dimension, for all $b$ and $z_\nu$. In this limit, $\mu_b=1/D$ is independent of the bitstring $b$ as well as of initial state and the measurement angle $\alpha$. For generic $\alpha\neq \pi/2$, the PoP continues to be Porter-Thomas distributed but with the appropriate mean given in Eq.~\ref{mu formula}.

\begin{figure}[]
\centering
\includegraphics[width=\linewidth]{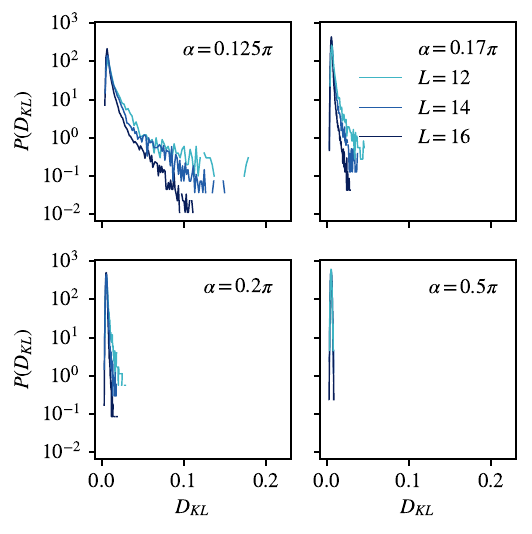}
\caption{The distribution of KL divergence between the Porter-Thomas distribution and PoP ($D_{KL}(\PoP_b||PT_b)$) across all $b$ (see Eq.~\ref{KL div eq} for definition). Different panels correspond to different $\alpha$ and in all cases the distributions get more concentrated around 0 with increasing $L$. Number of bins to obtain the histogram is kept same across all system sizes. The $PoP_b$ is obtained in a temporal ensemble with $10000$ states sampled with equal spacing in time interval $8000-10000$ in unit of $W^{-1}$.}
\label{fig:KL_div_distribution}
\end{figure}

While the results in Fig.~\ref{fig:single_bitstring_ditribution_temporal} show the Porter-Thomas distribution for the PoP for two representative bitstrings, to ascertain the agreement between $PT_b$ and $\PoP_b$ for arbitrary $b$, we look at their difference quantified by the Kullback–Leibler (KL) divergence.
For a particular bitstring $b$, the KL divergence between $PT_b$ and $\PoP_b$, is defined as 
\begin{align}
\label{KL div eq}
D_{KL}(\PoP_b||PT_b) =\int dp~ \PoP_b(p)  \ln \left(\frac{\PoP_b(p)}{PT_b(p)}\right)\,.
\end{align}
In the results presented, the integral in Eq.~\ref{KL div eq} is calculated by discretization and using the estimated $\PoP_b$ obtained by binning. 

In Fig.~\ref{fig:KL_div_distribution}, we present the distribution over $b$ of $D_{KL}(\PoP_b||PT_b)$ for different measurement angles $\alpha$.
The distributions are strongly concentrated near 0. This indicates that for most measurement bitstrings, the $\PoP_b(p)$ closely matches with $PT_b(p)$. 
For the smaller values of $\alpha$, close to zero, the distributions of $D_{KL}(\PoP_b||PT_b)$  develop a tail indicating deviations of the $\PoP_b$ from the $PT_b$ for some $b$ -- this can also be seen in Fig.~\ref{fig:single_bitstring_ditribution_temporal} for the smallest values of $\alpha$ presented. The finite-time effect behind the appearance of the tail at small $\alpha$ can be ruled out as the temporal ensemble is constructed with states sampled at sufficiently late-time (see Appendix~\ref{appendix:random phase ensemble}). However, these tails are suppressed with system size suggesting in the thermodynamic limit, $\PoP_b$ is indistinguishable from $PT_b$ for all $b$ except when $\alpha$ close to $0$.  

\begin{figure}
\centering
\includegraphics[width=\linewidth]{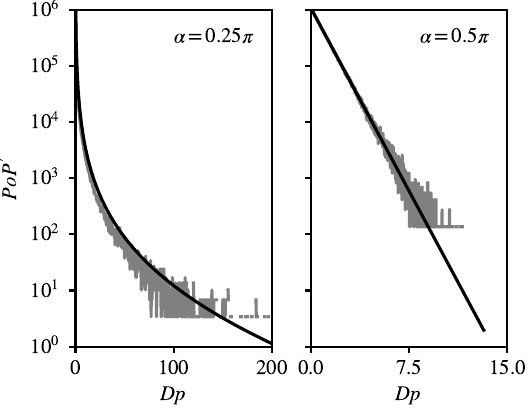}
\caption{ The $\PoP'$ distribution of measurement bitstrings for a single state at $Wt=10000$ following time evolution of an initial random direct product state under the $\ell$-bit Hamiltonian in Eq.~\ref{l-bit Hamiltonian}. The orange line denotes the sum of Porter-Thomas distributions shown in Eq.~\ref{single instatnt PoP}. Data for $L=20$. For $\alpha=\pi/2$ (right panel), all the Porter-Thomas distributions are with the same mean $1/D$ such that the $\PoP^\prime$ is also a Porter-Thomas distribution.
\label{fig:time_10000_L_20_all_bistring_dist}}
\end{figure}

As another evidence for the $\PoP_b$ being indistinguishable from $PT_b$, we also look at the distribution $PoP'$ of probabilities of bitstring outcomes across all bitstrings but calculated for a single state in the temporal ensemble. In other words, this is the distribution of $p(b) = |\braket{\psi_t|b}|^2$ computed for a fixed $\ket{\psi_t}$ across all $b$. 
Figure~\ref{fig:time_10000_L_20_all_bistring_dist} shows this distribution (gray lines) for two different values of $\alpha$, $\pi/4$ and $\pi/2$. For both these cases, the distribution fits to a sum of Porter-Thomas distributions corresponding to individual bitstrings (with means $\mu(b)$ given by Eq.~\ref{mu formula}),
\begin{align}
\label{single instatnt PoP}
\PoP'(p)=\frac{1}{2^L}\sum_b PT_b(p)\,.
\end{align}
The black lines in Fig.~\ref{fig:time_10000_L_20_all_bistring_dist} show these fits.
Note that for $\alpha=\pi/2$, $\mu_b=1/D~\forall~b$ such that $\PoP'$ is also a Porter-Thomas distribution with mean $1/D~$\cite{Manju2024}; this is evident in the right panel of Fig.~\ref{fig:time_10000_L_20_all_bistring_dist}. 

It is known that for general bitstring measurements, the PoPs are expected to be Porter-Thomas distributed with a mean of $1/D$ for random states~\cite{shaw2024universal}. Such random states are characterized by a Page value for the von Neumann entropy of bipartite entanglement~\cite{page1993average}. On the other hand, for $\alpha=\pi/2$, ($\sigma^x$-measurements) we find the PoPs to be a Porter-Thomas distribution with mean $1/D$ even though the states have much lower bipartite entanglement entropy than the Page value~\cite{bardarson2012unbounded,serbyn2013universal} indicating that they are far from a random state. 

\section{Matrix elements of Projected Ensemble and Scrooge Ensemble for $\ell$-bit Hamiltonian \label{Main: matrix_el_PE_Sc}}

In this section, we use the results for the PoP obtained in Sec.~\ref{sec:temp-ens-pop} to show that the projected ensemble approaches the Scrooge ensemble in the limit of $L_B,t\to\infty$. In Sec.~\ref{Matrix elements of Projected ensemble} we derive the expressions for the matrix elements of the $k^{\rm th}$ moments of the PE and in Sec.~\ref{Matrix elements of Scrooge ensemble} we derive the same for the Scrooge ensemble, and show that they are identical.

Before proceeding, it will be useful to define some notation. 
The set of bitstrings corresponding to the $\sigma^z$-product states in subsystem $A$, $B$, and the entire system are denoted by ${\cal Z}_A$, ${\cal Z}_B$, and ${\cal Z}_{AB}$ respectively. The bitstrings in $A$ are denoted as $z_A\equiv (z_{A,1},z_{A,2},\dots,z_{A,L_A}) \in {\cal Z}_A$ where $z_{A,i}=0,1$ denotes the bit at site $i$.  Similarly, the bitstrings in $B$ are denoted as $z_B\equiv (z_{B,L_A+1},z_{B,L_A+2},\dots,z_{B,L})\in {\cal Z}_B$. The bitstrings in the entire system are the concatenation of the bitstrings $z_A$ and $z_B$ such that $z_{AB} \equiv (z_{A},z_{B}) \in {\cal Z}_{AB}$.
Since we will be interested in arbitrary moments of the PE and the Scrooge ensemble, we denote the basis states in the $k$-replicated Hilbert space of $A$ as 
\eq{
\ket{Z_A} \equiv \ket{z_{A}^{(1)},z_{A}^{(2)},\cdots,z_{A}^{(k)}} = \otimes_{i=1}^k \ket{z_{A}^{(i)}}\,.
\label{eq:basis-HA-k-replica}
}
Finally, for a given initial state of the form Eq.~\ref{initial direct product state}, parametrized by $({\bm \theta},{\bm \phi})$ we define tensors $[M]_{i,s}$ on each site $i$ with
\eq{
M_{i,0}=\cos(\theta_i/2)\,,~~M_{i,1}=e^{\imath\phi_i}\sin(\theta_i/2)\,,
}
and we also define
\eq{
M_{z_{AB}} = M_{z_A}M_{z_B} = \left(\prod_{i=1}^{L_A} M_{i,z_{A,i}}\right)\left(\smashoperator[r]{\prod_{j=L_A+1}^L}M_{j,z_{B,j}}\right)\,.
}
With this notation, the initial state (Eq.~\ref{initial direct product state}) can be rewritten as 
\eq{
\ket{\psi_0} = \sum_{z_{AB}\in {\cal Z}_{AB}}M_{z_{AB}}\ket{z_{AB}}\,,
}
and the reduced density matrix of $A$ at infinite-time (Eq.~\ref{eq:rhoA-inf}) can be rewritten as 
\eq{
\rho_{A,\infty} = \sum_{z_A \in {\cal Z}_A}p(z_A)\ket{z_A}\bra{z_A}\,;~~p(z_A) = |M_{z_A}|^2\,.
\label{eq:pZA-def}
}

\subsection{Matrix elements of the projected ensemble}
\label{Matrix elements of Projected ensemble}
We will first express the matrix elements of the $k^{\rm th}$ moment of the PE at an arbitrary time $t$ and an arbitrary measurement basis $(\alpha)$ on $B$ and then take the limit of $t\to\infty$ for which we will average the matrix elements over the temporal ensemble defined in Sec.~\ref{sec:temp-ens-pop}. 
Since the states $\{\ket{z_{AB}}\}$ are also the eigenstates of the $\ell$-bit Hamiltonian (Eq.~\ref{l-bit Hamiltonian}), with eigenvalues $\{E_{z_{AB}}\}$, the state at time $t$ can be written as
\eq{
\ket{\psi_t} = \sum_{z_{AB}\in {\cal Z}_{AB}}e^{-\imath tE_{z_{AB}}}M_{z_{AB}}\ket{z_{AB}}\,.
\label{eq:psit}
}
If the measurement on $B$ produces a bitstring $b$, then the {\it unnormalized} conditional state on $A$ is given by
\eq{
\ket{\tilde{\psi}_{t,A}(b)} = \sum_{z_A\in {\cal Z}_A}M_{z_A}\varphi_{t}(b|z_A)\ket{z_A}\,,
\label{eq:psi-tilde}
}
where 
\eq{
\varphi_t(b|z_A) = \sum_{z_B\in {\cal Z}_B}M_{z_B}e^{-\imath t E_{z_{AB}}}\braket{b|z_B}\,,
\label{eq:varphi}
}
and the normalized version of the state in Eq.~\ref{eq:psi-tilde} is given by
\eq{
\ket{{\psi}_{t,A}(b)}=\frac{\ket{\tilde{\psi}_{t,A}(b)} }{\sqrt{p_t(b)}}\,;~p_t(b) = \braket{\tilde{\psi}_{t,A}(b)|\tilde{\psi}_{t,A}(b)}\,.
\label{eq:psit-b}
}
The quantity $p_t(b|z_A)\equiv|\varphi_t(b|z_A)|^2$ defined via Eq.~\ref{eq:varphi} has the physical meaning that it is the probability of obtaining the bitstring $b$ upon measurements (at angle $\alpha$) on $B$ at time $t$ given the $\sigma^z$-bitstring in $A$ is $z_A$.
The probability of obtaining the bitstring $b$ in $B$ due to the measurement on the state $\ket{\psi_t}$ in Eq.~\ref{eq:psit} is therefore
\eq{
p_t(b) = \smashoperator[r]{\sum_{z_A\in {\cal Z}_A}}p_t(b|z_A)p(z_A)=\smashoperator[r]{\sum_{z_A\in {\cal Z}_A}}|\varphi_t(b|z_A)|^2p(z_A)\,,
\label{eq:ptb}
}
with $p(z_A)$, given in Eq.~\ref{eq:pZA-def}, the probability of measuring the bitstring $z_A$ in $A$ which remains constant with $t$.
Using Eq.~\ref{eq:psit-b} and Eq.~\ref{eq:ptb}, an arbitrary matrix element of the $k^{\rm th}$ moment of the PE (defined in Eq.~\ref{moment definition}) can be written as 
\eq{
\begin{split}
\braket{Z^\prime_A|\rho_{A,t}^{(k)}|Z_A} 
=\prod_{i=1}^{k}&M^\ast_{z^{\prime(i)}_{A}}M_{z^{(i)}_{A}}\times\\
&\sum_b \frac{\prod_{i=1}^{k}\varphi_t^\ast(b|z^{\prime(i)}_{A})\varphi_t(b|z^{(i)}_{A})}{[p_t(b)]^{k-1}}\,.
\label{eq:rho-A-k-matel}
\end{split}
}
To estimate the matrix elements at late times, it is useful to introduce the $k^{\rm th}$ {\it no-resonance condition}~\cite{mark2024maximum}. 
The spectrum of a Hamiltonian satisfies satisfies this condition if for any two different size-$k$ sets of energy eigenvalues, their sums are different. 
We expect the $\ell$-bit Hamiltonian to satisfy this condition due to the random interactions between the spins. Under this assumption and noting the time-dependence of $\varphi_t$ in Eq.~\ref{eq:varphi}, we expect any matrix element in Eq.~\ref{eq:rho-A-k-matel} to have an oscillatory component and therefore evaluate to zero, unless $\ket{Z_A}$ and $\ket{Z^\prime_A}$ are equal up to a permutation, $(z_A^{(1)},\cdots,z_A^{(k)}) = \sigma(z_A^{\prime(1)},\cdots,z_A^{\prime(k)})$ for some permutation operator $\sigma\in S_k$ with $S_k$ the permutation group of $k$ elements.  
More over note that the matrix elements are invariant under the permutations of the replicas.
The $k^{\rm th}$ moment of the PE can therefore be expressed as 
\eq{
\rho_A^{(k)} = \sum_{Z_A}\sum_{\sigma\in S_k}\ket{Z_A}\bra{\sigma(Z_A)}\varrho^{(k)}_{{Z_A}}\,,
\label{eq:rhoA-perm}
}
where $|Z_A|$ is the number of unique $z_A$ bitstrings among the $k$-replicas, and the matrix element is given by
\eq{
\varrho^{(k)}_{{Z_A}} =  \sum\limits_{b} \frac{\prod_{i=1}^k p(z_A^{(i)}) p_t(b | z_A^{(i)})}{[\sum_{z^A\in {\cal Z}^A} p(z_A) p_t(b | z_A)]^{k-1}}\,.
\label{eq:rho-A-k-diag}
}
We will now use the observation made in Sec.~\ref{sec:temp-ens-pop} that the matrix elements of $\rho_{A,t}^{(k)}$ saturate to steady-state values at long times with temporal fluctuations that decay to zero with system size (see Fig.~\ref{fig:timeevo_single_state_alpha_0p25_mom2}).
This allows us to approximate the matrix elements in Eq.~\ref{eq:rho-A-k-matel} at late-times as their temporal average which can be implemented by averaging the matrix elements over the temporal ensemble, $\mathcal{E}_{\text{temp}}$, defined in Eq.~\ref{eq:TE-def}.

Before averaging the matrix element over the temporal ensemble, we note two points about the distribution of $p_t(b|z_A)$ in $\mathcal{E}_{\text{temp}}$.
Firstly, $\varphi_t(b|z_A)$ in Eq.~\ref{eq:varphi} is a sum over $z_B$ of phases $e^{\imath E_{z_{AB}}t}$ that depend on $z_A$, resulting in $p_t(b|z_A)=|\varphi_t(b|z_A)|^2$ for different $z_A$ being uncorrelated in the temporal ensemble at late times except when $\alpha\to0$. In the latter case, $p_t(b|z_A)$ is independent of $z_A$ and causes, large deviation of the point at $\alpha=0$ in Fig.~\ref{fig:L_alpha_dependence_mat_elements} from the Scrooge ensemble arises from this correlation. 
Secondly, $\varphi_t(b|z_A)$ can be interpreted as the amplitude of a bitstring $b$ in a time-evolved state with a modified $\ell$-bit Hamiltonian $H_B$ which is obtained from $H$ (Eq.~\ref{l-bit Hamiltonian}) by setting the degrees of freedom in $A$ to classical values determined by $z_A$.
We therefore expect the probability distribution of $p_t(b|z_A)$ to be described by a Porter-Thomas distribution \eqref{eq:PT-def} (as discussed in Sec.~\ref{sec:temp-ens-pop}) with a mean given by
\begin{equation}
\mu(b) = \sum_{z^B \in {\cal Z}^B}|M_{z_B}|^2 |\langle b|z_B \rangle|^2\,.
\end{equation}
The matrix element in Eq.~\ref{eq:rho-A-k-diag} upon averaging over the temporal ensemble can thus be expressed as 
\eq{
\varrho_{Z_A}^{(k)}=\sum\limits_{b}\int\bigg(\prod_{z_A}&\frac{dp(b|z_A)}{\mu(b)}e^{-\frac{p(b|z_A)}{\mu(b)}}\bigg)\times \nonumber\\ &\frac{\prod_{i=1}^k p(z_A^{(i)}) p_t(b | z_A^{(i)})}{[\sum_{z^A\in {\cal Z}^A} p(z_A) p_t(b | z_A)]^{k-1}}\,.
}
Though $p(b\vert z_A{A}) \in [0,1]$, since $\mu(b)$ is typically much smaller than $1$ and the integrand decays exponentially with $p(b\vert z_A)/\mu(b)$, we can extend the upper limit of integration over $p(b\vert z^{A})$ from 1 to $\infty$. Performing a variable transformation $\frac{p(b|z_A)}{\mu(b)}p(z_A) \rightarrow Y_{z_A} $ and noting that $\sum_{b} \mu(b)=1$ , we obtain for the matrix element
\begin{align}
\label{eq:PE-matel-final}
\varrho_{Z_A}^{(k)} = \int \left(\prod_{z_A}\frac{dY_{z_A}}{p(z_A)}e^{-\frac{Y_{z_A}}{p(z_A)}}  \right)\frac{\prod\limits_{i=1}^k Y_{z_A^{(i)}}}{\left(\sum\limits_{z_A} Y_{z_A}\right)^{k-1}}\,. 
\end{align}
We will next compute the corresponding matrix elements of the Scrooge ensemble and show that they are identical to the result obtained above  in Eq.~\ref{eq:PE-matel-final}.

\subsection{Matrix elements of the Scrooge ensemble}
\label{Matrix elements of Scrooge ensemble}

In order to derive the matrix elements of the Scrooge ensemble, corresponding to $\rho_{A,\infty}$, we will use the fact that it can be viewed as a distortion of the Haar ensemble as in Eq.~\ref{eq:Scrooge-def} and its $k^{\rm th}$ moment, following Eq.~\ref{rho k first expression} is given by
\eq{
\rho^{(k)}_{A,{\rm Scr}}=D_A\int d\psi~P_{\text{Haar}}(\psi) \frac{(\sqrt{\rho_{A,\infty}}|\psi \rangle \langle \psi|\sqrt{\rho_{A,\infty}})^{\otimes k} }{\braket{\psi|\rho_{A,\infty}|\psi}^{k-1}} \,,
}
where $D_A = 2^{L_A}$ is the Hilbert-space dimension of $A$ and the integral is over Haar-random states in $A$. Note from Eq.~\ref{eq:pZA-def} that $\rho_{A,\infty}$ is diagonal in the $\sigma^z$-product state basis with eigenvalues $p(z_A)$, such that the matrix element can be written as
\eq{
\braket{Z_A^\prime|\rho_{A,{\rm Scr}}^{(k)}|Z_A}=
D_A\int &d\psi~P_{\rm Haar}(\psi)\times\nonumber\\
&\frac{\prod_{i=1}^k \sqrt{p(z_A^{(i)})p(z_A^{\prime(i)})}\psi_{z_A^{(i)}}\psi^*_{z_A^{\prime(i)}}}{[\sum_{z_A}p(z_A)|\psi_{z_A}|^2]^{k-1}}\,.
}
where $\psi_{z_A} = \braket{z_A|\psi}$.
Due to the Haar averaging the product of the amplitudes $\psi_{z_A^{(i)}}\psi_{z_A^{\prime(i)}}$ in the numerator of the above equation, the matrix elements vanishes unless $\ket{Z_A}$ and $\ket{Z_A^\prime}$ are equal upto a permutation, exactly analogous to the PE at late times. The $k^{\rm th}$ moment of the Scrooge ensemble therefore has the form identical to that in Eq.~\ref{eq:rhoA-perm} but with the corresponding matrix element given by
\eq{
\varrho^{(k)}_{{\rm Scr},Z_A}=
D_A\int &d\psi~P_{\rm Haar}(\psi)
&\frac{\prod_{i=1}^k p(z_A^{(i)})|\psi_{z_A^{(i)}}|^2}{[\sum_{z_A}p(z_A)|\psi_{z_A}|^2]^{k-1}}\,.
\label{eq:scrooge-before-haar-integration}
}
The integral over the Haar random states can be performed using standard techniques (see Appendix~\ref{appendix:Scrooge} for details) which yields
\eq{
 \varrho^{(k)}_{{\rm Scr},Z_A} = \int \left(\prod_{z_A}\frac{dX_{z_A}}{p(z_A)}e^{-\frac{X_{z_A}}{p(z_A)}}  \right)\frac{\prod\limits_{i=1}^k X_{z_A^{(i)}}}{\left(\sum\limits_{z_A} X_{z_A}\right)^{k-1}}\,,
 \label{eq:scrooge-matel-final}
}
for the matrix element $\varrho^{(k)}_{{\rm Scr},Z_A}$. Comparing the expression for the Scrooge ensemble in Eq.~\ref{eq:scrooge-matel-final} with that obtained for the PE in Eq.~\ref{eq:PE-matel-final}, it is straightforward to see that they are identical with the identification of the integration variables $X_{z_A} \leftrightarrow Y_{z_A}$. This concludes our proof of the statement that if $\PoP$ of bitstrings approach uncorrelated PT distributions, the PE at late times approaches the Scrooge ensemble corresponding to $\rho_{A,\infty}$.

\section{Conclusion}
\label{Conclusion}
In this work, we investigated the projected ensemble obtained by single site measurements of all qubits outside a small subsystem of a unitarily evolving system
{with has (quasi)locally conserved charges. In particular, we numerically studied a strongly disordered Floquet spin-chain known to exhibit MBL dynamics for very long timescales as well as an $\ell$-bit model, a phenomenological model which is manifestly MBL and has an extensive number of $1$-local conserved charges, $\{\sigma^z_i\}$. Our analysis suggests that for generic measurement bases not very close to the conserved charges, the PE converges to the Scrooge ensemble (in the limit of large number of measured qubits)}.
Specifically for the $\ell$-bit model, the projected ensemble converges to a Scrooge ensemble if the qubits are measured along angles ($\alpha$ that parametrize the polar angle of the direction $\hat{n}$ for the Pauli observables ${\bm \sigma}\cdot\hat{n}$ that are measured on each site) that are not close to the direction of the conserved charges ($\alpha{=}0$ for $\sigma^z_i$). Numerical tests on first three moments confirm that the the PE indeed approaches a $k{=}3$-design of the Scrooge ensemble. For $\alpha$ near $0$, the bitstring measurement probabilities develop correlations that cause a deviation from Scrooge ensemble moments. It is unclear how these correlations scale with system size. Numerical results within available system sizes show exponential decay of trace distances of the moments from the Scrooge values for all $\alpha\gtrsim 0.1\pi$.

This behavior is in contrast to cases with global conserved charges where the PE converges to a angle dependent convex mixture of Scrooge ensembles, called the generalized Scrooge ensemble~\cite{mark2024maximum,chang2024deep}. In this case, revelation of partial information about the charges in subsystem $B$ due to measurements also reveals partial information about the charges in subsystem $A$ as the two are related via the global constraint. 

On the other hand, in the case with locally conserved charges, even though measurements reveal information (even if partial) about the charges local to $B$, they do not reveal anything about the charges in the subsystem $A$. The PE should then approach the ensemble that maximizes the entropy subject to the constraints of the charges in $A$; this is nothing but the Scrooge ensemble constructed out of $\rho_{A,\infty}$. In the pathological case of $\alpha=0$, when the measurements basis perfectly coincides with the conserved charges and the asymptotic PE is different from the Scrooge ensemble. The asymptotic PE of states in $A$ obtained by $\sigma^z$-measurements on $B$ has a trivial form (see Appendix~\ref{app:z-meas}).

The convergence to the Scrooge ensemble can be related to the fact that the probability distribution $\PoP$ of the expectation values of $\Pi_{b}$ (projectors into specific bitstring states for the measured observables) in the temporal ensemble (made of states sampled from the time evolution trajectory) approach the Porter-Thomas distribution in the long time limit. We find, numerically, that the distribution indeed approaches the Porter-Thomas distribution. This empirical result, which is expected in the $\PoP$ of chaotically evolving systems, is surprising here given that the $\ell$-bit Hamiltonian has a large number of local conserved charges. Though, we have not been able to prove these empirical results, we believe that this may be related to ergodicity in the unitary group $U(2^L)$ of the trajectories of all unitaries of the form $U_xU_t$ where $U_t$ is the unitary time evolution operator and $U_x$ are all elements of the stabilizer group of the bitstring states.

Though the results presented are for the $\ell$-bit Hamiltonian with finite interaction range (taken to be $\xi=2$ here), the results regarding the limiting distributions are valid even for longer range interacting systems with the same 1-local-conserved charges. The latter will approach the same limiting ensembles faster. A question for the immediate future is to understand the the mechanism that produces the power-law in time behavior in the approach to the Scrooge ensemble. In particular, it will be interesting to understand if this power-law decay is related to the power-law decay of temporal fluctuations of local observables in the MBL phase~\cite{serbyn2014quantum} or the power-law decay of bipartite purity (equivalent to logarithmic growth of bipartite entanglement)~\cite{bardarson2012unbounded,serbyn2013universal} in MBL systems. 

Finally, note that while the local conserved charges in the $\ell$-bit Hamiltonian were manifestly built-in, in a genuine MBL systems, they are expected to be emergent at strong disorder. It remains a question as to whether the PE and its higher moments can shed deeper insights and lead to better understanding of the MBL regime that goes beyond the breakdown of conventional ETH.

\begin{acknowledgments}
SR thanks P. W. Claeys and S. Mandal for fruitful discussions. GJS thanks M. Banik and A. Lakshminarayan for useful discussions. We thank the National Supercomputing Mission (NSM) for providing computing resources of ``PARAM Brahma'' at IISER Pune, which is implemented by C-DAC and supported by the Ministry of Electronics and Information Technology (MeitY) and Department of Science and Technology (DST), Government of India. SM thanks ICTS-TIFR for their hospitality during a research stay, during which this work was conceived and started.
SR acknowledges support from SERB-DST, Government of India under Grant No. SRG/2023/000858, from the Department of Atomic Energy, Government of India, under Project No. RTI4001, and from a Max Planck Partner Group grant between ICTS-TIFR, Bengaluru and MPIPKS, Dresden.
GJS acknowledges support from IHUB Quantum Technology Foundation, IISER Pune. The authors benefited from discussion during the meeting - Quantum Many-Body Physics in the Age of Quantum Information (code: ICTS/QMBPQI2024/11) organized by ICTS. 
\end{acknowledgments}

\appendix

\begin{widetext}

\section{Calculation of matrix elements for $k$th moments of the Scrooge ensemble\label{appendix:Scrooge}}
In this appendix, we provide some details of the derivation of the matrix elements of the moments of the Scrooge ensemble, in particular, the steps involved in obtaining the result in Eq.~\ref{eq:scrooge-matel-final} from Eq.~\ref{eq:scrooge-before-haar-integration}. The steps follow closely the derivation of the Scrooge ensemble in Ref.~\cite{mark2024maximum}. 

The integration over the Haar measure can be explicitly performed using 
\eq{
P_{\rm Haar}(\psi)d\psi = \frac{(D_A-1)!}{\pi^{D_A}}\delta(1-\sum_{z_A}|\psi_{z_A}|^2)\prod_{z_A}d^2\psi_{z_A}\,.
\label{eq:haar-dist}
}
However, since the integrand in Eq.~\ref{eq:scrooge-before-haar-integration} depends only on  $\{|\psi_{z_A}|^2\}$ the integral over the phases in $\psi_{z_A}$ can be done trivially. Defining a new variable $X_{z_A} = p(z_A)|\psi_{z_A}|^2$, Eq.~\ref{eq:scrooge-before-haar-integration} on using Eq.~\ref{eq:haar-dist} takes the form 
\eq{
\varrho^{(k)}_{{\rm Scr},Z_A}=
(D_A!)\int \left(\prod_{z_A}\frac{dX_{z_A}}{p(z_A)}\right)\delta\left(1-\sum_{z_A}\frac{X_{z_A}}{p(z_A)}\right)
\frac{\prod_{i=1}^k X_{z_A^{(i)}}}{[\sum_{z_A}X_{z_A}]^{k-1}}\,.
\label{eq:scrooge-before-haar-integration-app}
}
The $\delta$-function can be removed using a Laplace transform trick. We define 
\eq{
{f}^{(k)}_{Z_A}(t) = (D_A!)\int \left(\prod_{z_A}\frac{dX_{z_A}}{p(z_A)}\right)\delta\left(t-\sum_{z_A}\frac{X_{z_A}}{p(z_A)}\right)
\frac{\prod_{i=1}^k X_{z_A^{(i)}}}{[\sum_{z_A}X_{z_A}]^{k-1}}\,,
\label{eq:f-defn}
}
with the identification that $\varrho^{(k)}_{{\rm Scr},Z_A} = f^{(k)}_{Z_A}(t=1)$. The Laplace transform of Eq.~\ref{eq:f-defn} gives
\eq{
\tilde{f}^{(k)}_{Z_A}(s) = \int_0^\infty dt~e^{-st}{f}^{(k)}_{Z_A}(t)&=(D_A!)\int \left(\prod_{z_A}\frac{dX_{z_A}}{p(z_A)}\right)
\frac{\prod_{i=1}^k X_{z_A^{(i)}}}{[\sum_{z_A}X_{z_A}]^{k-1}}
\exp\left(-s\sum_{z_A}\frac{X_{z_A}}{p(z_A)}\right)\\
&=\frac{(D_A!)}{s^{D_A+1}}\int \left(\prod_{z_A}\frac{dX_{z_A}}{p(z_A)}e^{-\frac{X_{z_A}}{p(z_A)}}\right)
\frac{\prod_{i=1}^k X_{z_A^{(i)}}}{[\sum_{z_A}X_{z_A}]^{k-1}}\,,
}
where in the second line we performed a change of variables $X_{z_A}\to X_{z_A}/s$.
Since the inverse Laplace transform of $D_A!/s^{D_A+1}$ is simply $t^{D_A}$, we have
\eq{
\varrho^{(k)}_{{\rm Scr},Z_A} = f^{(k)}_{Z_A}(t=1) = \int \left(\prod_{z_A}\frac{dX_{z_A}}{p(z_A))}e^{-\frac{X_{z_A}}{p(z_A)}}\right)
\frac{\prod_{i=1}^k X_{z_A^{(i)}}}{[\sum_{z_A}X_{z_A}]^{k-1}}\,,
}
which is exactly the result in Eq.~\ref{eq:scrooge-matel-final}.

\section{Comparison of $\mathcal{E}_{\text{temp}}$ with random phase ensemble\label{appendix:random phase ensemble}}
\begin{figure}[H]
\centering
\includegraphics[width=0.5\textwidth]{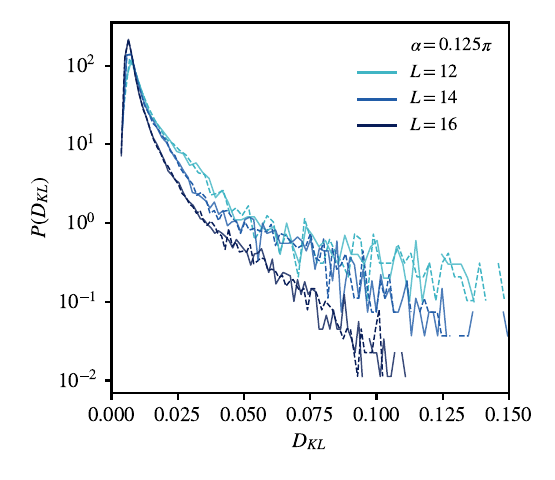} 
\caption{The random phase ensemble (dotted lines) is compared with the $\mathcal{E}_{\text{temp}}$ (solid lines) for $\alpha=\pi/8$. For all $L$ values, the distribution of $D_{KL}$ with PT for both ensembles match closely.\label{fig:random_phase_ensemble_KL_div_alpha_0p125}}
\end{figure}
Here, we further investigate the appearance of long tail in distribution of $D_{KL}$ for states in $\mathcal{E}_{\text{temp}}$. We construct the random phase ensemble ($\mathcal{E}_{\text{random-phase}}$)which is proposed to be the limiting ensemble of $\mathcal{E}_{\text{temp}}$ if the hamiltonian $H$ satisfies the $k$-th no-resonance condition and the initial state has finite overlap with all energy eigenstates\cite{mark2024maximum}. 
\begin{align}
\mathcal{E}_{\text{random-phase}}  =\left\{\sum_{j=1}^Dc_je^{-\imath \phi_j}\left\vert z_j\right\rangle\right\}.
\end{align}
where
$c_j=\langle \psi_0 | E_j \rangle$ where $E_j$ are energy eigenstates ( $\left|z_j\right \rangle$ for $\ell$-bit Hamiltonian) and $\phi_j$s are uniformly sampled from $[0,2\pi]$.
We keep equal number of states in $\mathcal{E}_{\text{random-phase}}$ and $\mathcal{E}_{\text{temp}}$. In Fig. \ref{fig:random_phase_ensemble_KL_div_alpha_0p125}, the distribution of $D_{KL}$ is almost identical for both the ensembles. This rules out any finite-time effect for the appearance of the log-tail. It can be attributed purely to the effect of measurement basis.

\section{Projected ensemble with $\sigma^z$-measurements \label{app:z-meas}}
In this section, we derive the $k{\rm th}$ moment of the PE for the case of $\sigma^z$-measurements ($\alpha=0$) and show that they are different from the Scrooge ensemble. Using the notation in Sec.~\ref{Main: matrix_el_PE_Sc}, the state at time $t$ is given by
\eq{
\ket{\psi_t} = \sum_{z_A\in {\cal Z}_A}\sum_{z_B\in {\cal Z}_B}M_{z_A}M_{z_B}e^{-\imath t E_{z_Az_B}}\ket{z_A z_B}\,,
}
such that a $\sigma^z$-measurement on $B$ yields the bitstring $z_B$ with probability $p(z_B) = |M_{z_B}|^2$ and the corresponding state on $A$ is
\eq{
\ket{\psi_{A,t}(z_B)} = \sum_{z_A\in {\cal Z}_A}M_{z_A}e^{-\imath t E_{z_Az_B}}\ket{z_A}\,.
}
Denoting $|M_{z_{A/B}}|^2 = p(z_{A/B})$, the $k^{\rm th}$ moment of the PE therefore has matrix elements
\eq{
\braket{Z_A^\prime|\rho_{A,t}^{(k)}|Z_A} = \sum_{z_B\in {\cal Z}_B}p(z_B) \prod_{i=1}^k M^\ast_{z_A^{\prime (i)}}M_{z_A^{(i)}} \exp\left[{-\imath t \left(E_{z_A^{(i)}z_B}-E_{z_A^{\prime(i)}z_B}\right)}\right]\,.
}
In the limit of $t\to\infty$, the $k^{\rm th}$ no-resonance condition again implies that the matrix element above is non-zero only if $\ket{Z_A^\prime}$ is a permutation of $\ket{Z_A}$ such that
\eq{
\rho_A^{(k)} = \sum_{Z_A}\sum_{\sigma\in S_k}\ket{Z_A}\bra{\sigma(Z_A)}\varrho^{(k)}_{{Z_A}}\,,
}
and the matrix element is given by
\eq{
\varrho^{(k)}_{{Z_A}} =  \prod_{i=1}^k p(z_A^{(i)})\,.
\label{eq:matel-z}
}
Note the structure of the $k^{\rm th}$ moment of the PE here is same as the Scrooge ensemble - location of the non-zero matrix elements are the same. However the matrix element for the Scrooge ensemble in Eq.~\ref{eq:scrooge-matel-final} can be written as
\eq{
\varrho^{(k)}_{{\rm Scr},Z_A}=\prod_{i=1}^k p(z_A^{(i)})~\times~\int\left(\prod_{z_A}d x_{z_A}~e^{-x_{z_A}}\right)\frac{\prod_{i=1}^k x_{z_A^{(i)}}}{(\sum_{z_A} x_{z_A}p(z_A))^{k-1}}\,.
\label{eq:matel-scr-app}
}
Note that the expressions in Eq.~\ref{eq:matel-z} and Eq.~\ref{eq:matel-scr-app} are obviously different in general from each other for general $\{p(z_A)\}_{z_A}$, which are set by the initial state, as the integral in the RHS of the above equation does not evaluate to unity in general.

\end{widetext}

\typeout{} 
\bibliography{ProjectedEnsemble.bib}	
\end{document}